\documentclass[ aps, prl, twocolumn, showpacs,nofootinbib, nobibnotes, superscriptaddress,longbibliography]{revtex4-2}
\usepackage{graphicx}
\usepackage{amsmath}
\usepackage{amssymb}
\usepackage{mathtools}
\usepackage{MnSymbol}
\usepackage{amsthm}
\usepackage{braket}
\usepackage{hyperref}
\usepackage{comment}
\usepackage{xcolor}

\def\abs#1{ {|#1|}}

\newcommand{\eps}{\epsilon}
\newcommand{\bs}[1]{\boldsymbol{#1}}
\newcommand{\Yb}{${}^{171}{\rm Yb}^+$}

\begin{document}
\title{Crosstalk Insensitive Trapped-Ion Entanglement through Coupling Matrix Engineering}
\author{Vikram Kashyap}
\email{vikk@umd.edu}
\affiliation{Joint Center for Quantum Information and Computer Science, University of Maryland - 20742, USA}
\affiliation{Department of Physics, University of Maryland - 20742, USA}
\affiliation{Department of Electrical and Computer Engineering, University of Washington - 98195, USA}
\author{Caleb Walton}
\author{Sara Mouradian}
\affiliation{Department of Electrical and Computer Engineering, University of Washington - 98195, USA}

\begin{abstract}
Optical crosstalk due to imperfect addressing in trapped-ion entangling gates generates unwanted non-local entanglement between target ions and their neighbors that is difficult to mitigate using standard quantum error correction. We present a method to design entangling operations that are inherently insensitive to crosstalk by engineering the effective qubit coupling matrix. By controlling the geometric phases generated in the motional modes of the ion string, we construct a coupling matrix that selectively excludes crosstalk-affected neighbor ions from the entangling operation. This approach requires no knowledge of the amount of crosstalk present and avoids the need for additional gate operations or modifications to the optical setup. We numerically demonstrate the construction of crosstalk-insensitive entangling pulses for target ion pairs within an equispaced 20-ion string and provide experimental validation of crosstalk-insensitive entanglement in a three-ion string.
\end{abstract}
\maketitle

Trapped-ion systems are a leading platform for quantum information processing with operation errors below the error correction threshold in small-scale systems~\cite{smith2025_singlequbit, loschnauer2024scalable}. However, imperfect optical addressing is an increasingly important source of infidelity. Optical crosstalk can, in principle, be directly minimized with careful optical engineering~\cite{shen2013,  chen2024_lowcrosstalk}. However, this is increasingly difficult for large-scale systems~\cite{Nigg2014Jun,Manovitz2022Mar,Pogorelov2021Jun}, especially as the field moves towards integrated optics~\cite{Mehta2016Dec,Mehta2020Oct,Niffenegger2020Oct, Shirao2022Jun,Binai-Motlagh2023Jul,sotirova2024}, where optical aberrations are more prominent. Optical crosstalk can also be directly canceled with equal-amplitude, opposite phase signals~\cite{flannery2024physical, clark2021_engineering}, though this requires precise calibration. Even in a system with no optical aberrations, reducing the beam waist of the addressing beam to reduce optical crosstalk will also introduce other errors that may be more difficult to correct: polarization becomes ill-defined at a tightly focused beam waist~\cite{Chen2002Feb,Maltsev2003Feb}, and errors will arise due to relative ion-beam motion, either due to beam pointing instability~\cite{Niffenegger2020Oct,chen2024_lowcrosstalk} or, more fundamentally, thermal motion of the ion~\cite{Li2020Dec,West2021Jan,Cetina2022Mar}. 

High fidelity single-qubit gates in the presence of optical crosstalk are possible using composite pulse sequences~\cite{merrill2014}
or quantum error correction~\cite{heusen2023}. However, optical crosstalk errors in entangling gates are more difficult to correct using error-correcting codes because unwanted entanglement is generated between nonlocal sets of qubits. Thus, mitigating crosstalk errors at the gate level is necessary~\cite{parrado_rodriguez2021, debroy2020, heusen2023}. 

Previous proposals for crosstalk-insensitive entangling operations use mid-gate single-qubit rotations to echo out unwanted entangling operations between target and neighbor ions~\cite{parrado_rodriguez2021, fang2022}. In this work, we instead directly construct entangling gates that are insensitive to crosstalk by exciting the vibrational modes of the ion string in ratios that produce an effective ion-ion coupling matrix that excludes interactions between the target ions and their crosstalk-affected neighbors.

We consider a 1-dimensional string of $N$ trapped ions with internal qubit degrees of freedom and global vibrational modes. The two-qubit entangling gates we design use one set of radial modes with frequencies $\nu_m$, Lamb-Dicke parameters $\eta_m$, and mode participation values $b_{m,j}$ for each mode $m$ and ion $j$. Our aim is to perform an entangling gate on two target ions $t_1$ and $t_2$ while leaving the remaining ions unaffected.
The ideal two-qubit entangling unitary is $\hat{U}_\text{ideal} = \exp\left(i \Theta \hat{\sigma}_{t_1}^{(\phi_{t_1})} \hat{\sigma}_{t_2}^{ (\phi_{t_2})}\right)$, where $\hat{\sigma}_{j}^{(\phi_j)}$ is a Pauli operator operating on ion $j$ and $\Theta$ is the desired rotation angle.
The operator $\hat{\sigma}_{j}^{(\phi_j)}=\sin(\phi_j) \sigma^{x}_j + \cos(\phi_j)\sigma^y_j$ has an axis of rotation set by the phase $\phi_j$. In a typical entangling gate, we use $\phi_{t_1}=\phi_{t_2}=\pi/2$ to produce an $XX$ operation, and set $\Theta = \pi/4$ to prepare a maximally entangled state from a $\ket{00}$ initial state.

Trapped-ion entangling gates based on the Mølmer-Sørensen gate use a laser pulse with a frequency spectrum that is symmetrically distributed around the qubit transition frequency $\omega_0$ and where the upper (``blue") and lower (``red") halves of the spectrum have phases $\phi_B$ and $\phi_R$ respectively.
The pulse is described by $f(\tau)$, the (normalized) spin-dependent force effected by the laser drive as a function of time, scaled such that $\max_\tau \abs{f(\tau)} = 1$.
The interaction picture Hamiltonian takes the general form~\cite{molmersorensen2000_entanglement, landsman2019}
\begin{align}
    \hat{H}_\text{int}(\tau) = & \frac{1}{2} \sum_j \Omega_j \sum_m b_{m,j} \eta_m \hat{\sigma}^{(\phi_j)} \hat{a}^\dagger_m e^{i \nu_m \tau} f(\tau) + \text{h.c.},
    \label{eq:Hint}
\end{align}
where $\hat{a}^\dagger_m$ is the raising operator for mode $m$.
The pertinent spin axis for the spin-dependent force on each ion is set by $\phi_j = [\phi_R+\phi_B]/2+\tilde{\phi}_j$, where the phase offset $\tilde{\phi}_j$ accounts for a possible path length difference of light traveling to ion $j$. We use the convention $\hbar=1$.

The spin-dependent force function $f(\tau)$ is designed to displace the motional modes in loops of size dependent on the spin of each illuminated ion, resulting in geometric phases attached to each spin configuration. In the simplest case, $f(\tau)$ is a sine-wave with frequency equal to the detuning of the laser drive from $\omega_0$ \cite{molmersorensen2000_entanglement}. The requirement that all $N$ modes are unexcited by the gate time $\tau_\text{gate}$ imposes $2N$ linear constraints on $f(\tau)$. By using parameterized pulses such as segmented AM pulses \cite{choi2014} or multitone pulses \cite{shapira2020}, it is possible to simultaneously satisfy all these constraints over any gate duration $\tau_\text{gate}$. With the mode closure requirements satisfied, the unitary implemented at time $\tau_\text{gate}$ takes the form of pairwise coupled qubit rotations ~\cite{choi2014, landsman2019}
\begin{align}
    \hat{U} = \prod_{j_1<j_2} \exp\left(i \theta_{j_1,j_2} \hat{\sigma}^{(\phi_{j_1})}_{j_1} \hat{\sigma}^{(\phi_{j_2})}_{j_2} \right),
    \label{eq:qubit_unitary}
\end{align}
of angles
\begin{align}
    \theta_{j_1,j_2} &= 2 \Omega_{j_1} \Omega_{j_2} J_{j_1,j_2},
    \label{eq:thetaij_def}
\end{align}
where $\bs{J}$ is a symmetric $N \times N$ matrix such that the coupling between qubits $j_1$ and $j_2$ is given by $2J_{j_1,j_2}$.

\begin{figure}
    \centering
    \includegraphics[width=1\linewidth]{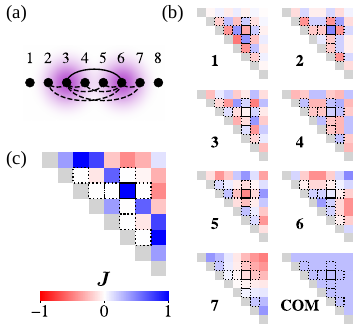}
    \caption{Crosstalk-insensitive qubit couplings. (a) A target gate (solid line) between two ions is driven using a laser (purple) incident on the two target ions. Spillover of the laser light onto ions neighboring the target ions results in crosstalk interactions (dashed lines). (b) The $\bs{J}^{(m)}$ single-mode coupling matrices of the radial modes of an equispaced $8$-ion string. The relative strengths of the target coupling (solid outline) and target-neighbor couplings (dashed outlines) depend on the mode. (c) A linear combination $\bs{J}$ of the single mode couplings matrices in (a) chosen such that the coupling of the target ions is nonzero but couplings between target and neighbor ions are zero.}
    \label{fig:crosstalk_and_mode_combination}
\end{figure}

The valid forms of the coupling matrix depend on the motional modes: $\bs{J}$ must be of the form $\bs{J} = \sum_m \chi_m \bs{J}^{(m)}$, where $\bs{J}^{(m)}$ is the single-mode coupling matrix associated with mode $m$: $J^{(m)}_{j_1,j_2} := b_{m,j_1,} b_{m,j_2}$. The single-mode coupling matrices for the radial vibrational modes of a string of $N=8$ equally-spaced ions (Fig.~\ref{fig:crosstalk_and_mode_combination}(a)) are displayed in Fig.~\ref{fig:crosstalk_and_mode_combination}(b). The vector $\bs{\chi}$ describing the linear combination of single-mode coupling matrices is determined by the geometric phases accrued in each mode, and is dependent only on the pulse function~\cite{choi2014, blumel2021_power}:
 \begin{align}
     \chi_m =- \frac{\eta_m^2}{4} \int_0^{\tau_\text{gate}} d\tau_1 \int_0^{\tau_1} d\tau_2 f(\tau_1)f(\tau_2) \sin(\nu_m [\tau_2-\tau_1]).
     \label{eq:chi_from_f}
 \end{align}

With perfect optical control we could achieve $\hat{U}=\hat{U}_\text{ideal}$ by setting $\Omega_{t_1/t_2} = \sqrt{\Theta/(2J_{t_1,t_2})}$ and $\Omega_j=0$ for all other ions. However, due to spillover of the controlling laser, each neighbor of the target ions will receive some fraction of the Rabi frequency $\Omega_{t_1/t_2}$ applied to the target ion.
Although this fraction may be different for each neighbor ion depending on its physical position, we can characterize the crosstalk by a maximal fraction $\eps$ such that $\Omega_n<\eps\Omega_{t_1/t_2}$ for all $n \in \mathcal{N}$, the set of neighbor ions. Furthermore the phase of the drive incident on the neighbor ions will depend on the path length of light to the ions, leading to an individual rotation axis $\phi_n$ for each $n \in \mathcal{N}$~\cite{fang2022}.

As per Eq.~\ref{eq:qubit_unitary}, nonzero intensity of the controlling laser on the neighbor ions leads to a target-neighbor crosstalk error described by a unitary that acts in addition to the target gate $\hat{U}_\text{ideal}$ and is given by
\begin{align}
    \hat{U}_\text{cross} = \sum_{\substack{t \in \mathcal{T} \\ n \in \mathcal{N}}} \exp\left(i \theta_{t,n} \hat{\sigma}^{(\phi_t)}_{t} \hat{\sigma}^{(\phi_n)}_{n} \right).
    \label{eq:error_unitary}
\end{align}
In Fig.~\ref{fig:crosstalk_and_mode_combination}(a) the target gate between ions 3 and 6 (solid line) is accompanied by 8 potential interactions with the four neighbor ions 2, 4, 5, and 7 (dotted lines). We presume that the profile of the controlling beam (purple) is tight enough not to overlap with next-nearest neighbors (ions 1 and 8) and so there are no interactions involving these ions.
Applying Eq.~\ref{eq:thetaij_def}, we can see that the crosstalk rotation angles are proportional to the target gate angle and depend on the relative strengths of the qubit couplings of the target and crosstalk pairs: $\theta_{t,n} = \eps \Theta J_{t,n}/J_{t_1,t_2}$. Because these angles are small, the worst-case infidelity is approximately given by~\cite{parrado_rodriguez2021}
\begin{align}
    \mathcal{I}_\text{cross} \approx \sum_{t \in \mathcal{T}, n \in \mathcal{N}} \theta_{t,n}^2 = \eps^2 \Theta^2 \sum_{t\in \mathcal{T}, n \in \mathcal{N}} \left(\frac{J_{t,n}}{J_{t_1,t_2}}\right)^2
    \label{eq:infidelity_approximation}
\end{align}

In this work we exploit the motional mode structure to construct gates that are inherently insensitive to crosstalk by engineering the coupling matrix $\bs{J}$. By producing a coupling matrix where $J_{t,n}=0$ for all target-neighbor pairs, we can ensure that $\theta_{t,n}=0$ independently of both the crosstalk fraction $\Omega_n/\Omega_{t_1/t_2}$ and the phase offset $\tilde{\phi}_n$ of the drive on each neighbor ion. We also desire that the target coupling $J_{t_1,t_2}$ be as large as possible to reduce the required laser power. 

Designing a qubit coupling matrix $\bs{J}$ through the preferential excitation of vibrational modes has been considered in the context of analog quantum simulation of spin models with uniform global drives~\cite{shapira2020, kyprianidis2024}. Since there are only $N$ degrees of freedom in $\bs{\chi}$ and $(N^2-N)/2$ parameters in $\bs{J}$, it is impossible to produce arbitrary inter-qubit couplings in this manner. To eliminate target-neighbor crosstalk, however, we only require control over $\abs{\mathcal{T}\times\mathcal{N}}+1 \leq 9$ degrees of freedom, a requirement that does not scale with $N$.

To consider a subset of qubit coupling values instead of the coupling matrix as a whole, we decompose each entry of the coupling matrix as
\begin{align}
    J_{j_1,j_2} = \bs{g}^{(j_1,j_2)} \cdot \bs{\chi}.
    \label{eq:J_gchi_decomp}
\end{align} The dependence of the coupling between ions $j_1$ and $j_2$ on their participation in each motional mode $m$ is described by $g^{(j_1,j_2)}_m := b_{m,j_1} b_{m,j_2}$. We seek a phase vector $\bs{\chi}$ that produces a qubit coupling matrix where $J_{t,n} = \bs{\chi} \cdot \bs{g}^{(t,n)} = 0$ for all target-neighbor pairs and where $J_{t_1,t_2}$ is as large as possible. When $\bs{\chi} \cdot \bs{g}^{(j_1,j_2)} = 0$, we say that this phase vector $\bs{\chi}$ ``eliminates" the coupling $J_{j_1,j_2}$.

Grouping the $g$-vectors describing crosstalk interactions into a set $\mathcal{R} = \left\{\bs{g}^{(t,n)} | t \in \mathcal{T}, n \in \mathcal{N}\right\}$ and defining the vector space $R:= \text{span}(\mathcal{R})$, we pick $\bs{\chi}$ in the orthogonal complement $R^\perp$ to ensure that there is zero target-neighbor coupling. The dimension of $R$ cannot be more than 8, the maximum number of target-neighbor pairings. Thus, as long as $N\geq 9$, the orthogonal space $R^\perp$ has positive dimension and there exists a $\bs{\chi}$ that eliminates all target-neighbor crosstalk interactions. Furthermore, in shorter strings a crosstalk-insensitive coupling matrix may still exist for a given pair of target ions because the target ions may have fewer than 4 neighbors and because the mirror symmetry of the string across the center may lead to $R$ having dimension less than the number of crosstalk interactions. An example of a crosstalk-insensitive coupling matrix for entangling target ions $\mathcal{T}=\{3,6\}$ in an 8-ion string is provided in Fig.~\ref{fig:crosstalk_and_mode_combination}(b). The crosstalk couplings (squares with dotted outline in Fig.~\ref{fig:crosstalk_and_mode_combination}(b)) for the 8 target-neighbor crosstalk interactions (dotted lines in Fig.~\ref{fig:crosstalk_and_mode_combination}(a)) are simultaneously eliminated by picking $\bs{\chi}=(-0.547, 0.412, 0.446, 0.028, -
1.306, -0.303, 0.678, 0.592)$. While the crosstalk-insensitive coupling matrix contains the couplings for all qubit pairs, the description of couplings using $g$-vectors allows us to concern ourselves with only controlling the target and crosstalk couplings.

To implement the target entangling operation, we require that $\bs{\chi} \in R^\perp$ be chosen such that $\theta_{t_1,t_2} = 2 \Omega_{t_1/t_2}^2 [ \bs{\chi} \cdot \bs{g}^{(t_1,t_2)}] = \Theta$. It is therefore necessary that $\bs{g}^{(t_1,t_2)}$ have nonzero projection onto $R^\perp$. We can quantify the independence of the target interaction from the crosstalk interactions using the ratio $a_{t_1,t_2} = \lvert\lvert \text{proj}_{R^\perp}(\bs{g}^{(t_1,t_2)}) \rvert\rvert/\lvert\lvert \bs{g}^{(t_1,t_2)} \rvert\rvert $, where $||\cdot||$ denotes the euclidean norm. If $a_{t_1,t_2}\neq0$, a valid coupling matrix exists for the set of motional modes under consideration where all crosstalk couplings are zero but the target coupling is nonzero. Furthermore, by Eq.~\ref{eq:J_gchi_decomp} we have the bound $\lvert J_{t_1,t_2} \rvert \leq a_{t_1,t_2}\lvert \lvert \bs{g}^{(t_1,t_2)}\rvert\rvert \lvert \lvert \bs{\chi} \rvert \rvert$ for all $\bs{\chi} \in R^\perp$. For a given gate time, and therefore a bounded value for $||\bs{\chi}||$, the inverse of $a_{t_1,t_2}$ is correlated with the peak Rabi frequency $\Omega_{t_1/t_2}$ required to implement the target gate.

Whether or not a crosstalk-insensitive coupling matrix exists therefore depends on the set of motional modes available and the particular pair of target ions being considered. We choose as an example the modes of ion strings with even ion spacing, which are commonly used in surface trap architectures and which have advantages in string stability, cooling, optical control, and measurement~\cite{lin2009, johanning2016}. Equispaced strings have modes that can be closely approximated by sinusoids \cite{kyprianidis2024, johanning2016}.
These sinusoidal modes have been shown to be capable of producing a variety of inter-qubit couplings useful for quantum simulation~\cite{kyprianidis2024}. In SM1, we prove that for this set of modes $a_{t_1,t_2}>1/\sqrt{8}$ for all pairs of target ions that do not include the two outermost ions ~\cite{supplemental_material}. This implies that a crosstalk-insensitive coupling matrix can be implemented in a power-efficient manner for all such target pairs of ions. For any other set of motional modes, the capacity for the modes to produce crosstalk-insensitive gates can be evaluated using the target-crosstalk independence parameters in the same manner.

To demonstrate the efficacy of our gate construction scheme, we design multi-tone pulses that implement crosstalk-insensitive entangling gates between pairs of ions in an equispaced 20-ion string. Following Ref.~\cite{shapira2020} we consider multi-tone pulses consisting of $M$ pairs of tones of the form $\omega_0\pm\omega_l$, each described by a detuning $\omega_l$ and a tunable amplitude $r_l$.  The phases of the higher and lower tones in each pair are $\phi_B$ and $\phi_R$ respectively. To ensure that $f(\tau)$ is normalized, we impose $\sum_l r_l=1$. Negative values of $r_l$ correspond physically to the change $\phi_{B/R} \rightarrow \phi_{B/R}+\pi$, but we absorb the resulting negative sign from $\hat{\sigma}^{(\phi_j)}_j \rightarrow -\hat{\sigma}^{(\phi_j)}_j$ into $r_l$. The multitone pulse function is given by
\begin{align}
    f_\text{mt}(t) = \sum_{l=1}^M r_l \cos(\omega_l \tau + [\phi_R-\phi_B]/2).
    \label{eq:multitone_pulse_func}
\end{align}
Further details of the derivation of the pulse function are presented in SM2 \cite{supplemental_material}. Applying the formula for $\bs{\chi}$ in Eq.~\ref{eq:chi_from_f} to the multitone pulse function in Eq.~\ref{eq:multitone_pulse_func}, we find a quadratic dependence of $\bs{\chi}$ on $\bs{r}$. This necessitates nonlinear optimization of the tone amplitudes to implement a desired qubit-coupling.

\begin{figure}
    \centering
    \includegraphics[width=1\linewidth]{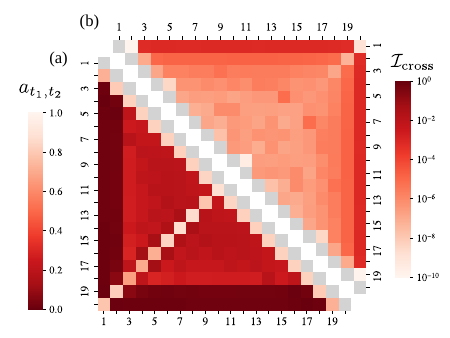}
    \caption{(a) The independence of the crosstalk and target gates, as indicated by the parameter $a_{t_1,t_2} = \lvert\lvert \text{proj}_{R^\perp}(\bs{g}^{(t_1,t_2)}) \rvert\rvert/\lvert\lvert \bs{g}^{(t_1,t_2)} \rvert\rvert $. (b) The infidelity $\mathcal{I}_\text{cross}$ due to target-neighbor crosstalk in a 20-ion equispaced string for multitone driving pulses optimized to eliminate target-neighbor crosstalk. Infidelity is calculated using Eq.~\ref{eq:infidelity_approximation} using crosstalk $\eps=0.02$ and target gate angle $\Theta=\pi/4$. Higher target-crosstalk independence correlates with lower crosstalk infidelity from the optimized pulse.}
    \label{fig:independence_and_pulse_infidelity}
\end{figure}

Using the differential evolution optimizer provided by \texttt{SciPy} \cite{scipy2020}, we optimize pulses for crosstalk-insensitivity in $\Theta=\pi/4$ entangling gates on each possible target pair of ions. For each pair we calculate the target-crosstalk independence parameter $a_{t_1,t_2}$ shown in Fig.~\ref{fig:independence_and_pulse_infidelity}(a).
Despite deviations of the modes of finite equispaced ion strings from the perfectly sinusoidal approximation, the target-crosstalk independence values are high for all target pairs in the bulk of the string, qualitatively matching the analytical results for sinusoidal modes.
The independence values give us confidence that efficient crosstalk-insensitive gates exist for all pairs in the bulk of the string. To find crosstalk-insensitive multi-tone entangling pulses, we optimize within reasonable experimental constraints for low infidelity (Eq.~\ref{eq:infidelity_approximation}). The infidelity due to crosstalk is presented in Fig.~\ref{fig:independence_and_pulse_infidelity}(b) for the crosstalk fraction $\eps=0.05$. While any target pair with $a_{t_1,t_2}>0$ indicates that a perfectly crosstalk-insensitive coupling matrix exists for the target gate, the target-neighbor infidelity obtained from the optimized pulses is not precisely zero because it may not be possible to implement a perfectly crosstalk-insensitive coupling with the specified gate time and maximum laser power. The pulses calculated for Fig.~\ref{fig:independence_and_pulse_infidelity}(b) all have duration $\tau_\text{gate}=250\mu\text{s}$, contain $M=50$ pairs of tones, use $\phi_B = \phi_R = 0$, and have a maximum Rabi frequency $\Omega_{t_1/t_2} \leq 2\pi \times 2.5 \text{MHz}$. The tones $\omega_l$ are chosen to be harmonics of the gate time, i.e. of the form $2\pi n/\tau_\text{gate}$ for integer n, and uniformly spread throughout the motional mode frequency range.

It is also possible to produce a desired $\bs{\chi}$ by using a series of adiabatic Mølmer-Sørensen operations \cite{molmersorensen2000_entanglement} each targeting a single mode to accrue the appropriate phase in each mode one after another. This method is prohibitively slow for long strings because it requires $N-1$ adiabatic operations to produce all the required relative phases between the modes and because the gate time must be long enough for the drive to couple to a single mode out of the $N$ modes. 

\begin{figure}
  \centering
  \includegraphics[width=\linewidth]{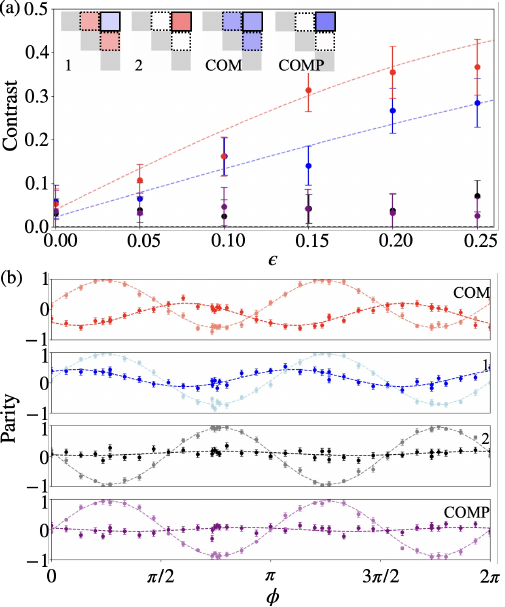}
  \caption{ a) Parity between the target and center ion for single mode excitation (black, blue, and red) and composite gate (purple) as a function of crosstalk. The inset shows the single-mode coupling matrices $\bs{J}^{(m)}$ for the three modes and the composite coupling matrix $\bs{J}$. b) The parity between the two target ions (pale) and target-center ions (dark) for each gate at $\eps=0.25$.}
  \label{fig:experiment}
\end{figure}

Nevertheless, it may be useful for short strings and we use this method to experimentally demonstrate crosstalk-insensitive gates by engineering a composite coupling matrix. We use the QSCOUT platform~\cite{clark2021_engineering, supplemental_material} to run entangling gates on the outer two ions of a three-ion chain while increasing the optical field strength on the center ion to artificially implement crosstalk. We use a series of adiabatic Mølmer-Sørensen operations~\cite{molmersorensen2000_entanglement} each targeting a single mode to accrue the appropriate phase in each mode one after another. The inset of Fig.~\ref{fig:experiment}(a) shows the single-mode coupling matrices $\bs{J}^{(m)}$ for each of the modes in a three-ion chain. As expected from the mirror symmetry of the chain, the ``tilt'' mode (mode 2) provides natural crosstalk immunity against the crosstalk interactions between ion pairs $(1,2)$ and $(2,3)$ because $g^{(1,2)}_2 = g^{(2,3)}_2 = 0$. In contrast, the center ion has strong participation in the other two modes.

To begin, we compare the fidelity of three single-mode Mølmer-Sørensen entangling operations.  The gates are run with a duration of 250\,$\mu$s, and a $-15$\,kHz detuning from the target radial mode. A parity measurement after a controlled rotation converts the off-diagonal coherence of the created Bell state into measurable population oscillations, and the oscillation contrast directly bounds entanglement~\cite{haljan2005entanglement, sackett2000experimental}. Fig.~\ref{fig:experiment}(a) shows the contrast of these parity measurements between the center and target ions, bounding the degree of unwanted entanglement generated. As expected, no entanglement is generated between the target and center ion when only mode 2 is excited (black). However, when using mode 1 (blue) and the COM mode (red) to mediate entanglement generation, the maximum parity increases significantly, matching theoretical expectations (dotted lines). Figs.~\ref{fig:experiment}(b-d) show the full parity curves for both target-neighbor and target-target pairs at $\eps=0.25$ verifying that the increase in target-neighbor entanglement is reflected in a decrease in target-target fidelity.

Utilizing only the tilt mode corresponds to the crosstalk-insensitive phase vector $\bs{\chi} \propto (0,1,0)$, but using our framework we note that the crosstalk-insensitive space for this gate also includes $\bs{\chi} \propto (1,0,1)$. These two values for $\bs{\chi}$ happen to indicate the same set of relative phases between the modes, which is why the single mode coupling matrix for mode 2 in the inset of Fig.~\ref{fig:experiment}(a) is simply the negative of the composite coupling matrix. However the manner of implementing these $\bs{\chi}$ vectors is distinct and so demonstrates the principles of coupling matrix engineering. We implement the phase vector $\bs{\chi}=(1,0,1)$ using a $\theta_{t_1,t_2} = \pi/6$ pulse utilizing mode 3 (COM) followed by a $\theta_{t_1,t_2}=\pi/12$ pulse utilizing mode 1, both with the same parameters as described above. The first pulse generates temporary entanglement between the center and target ions, but this is reversed by the second pulse. The low target-neighbor parity measurements for the composite gate (purple) in Fig.~\ref{fig:experiment}(a,b) demonstrate that the single-mode coupling matrices have been successfully combined to produce a crosstalk-insensitive coupling.

In this manuscript, we have presented a method to design crosstalk-insensitive entangling gates, eliminating a significant source of coherent error in trapped-ion devices. Other sources of coherent error, such as coupling to the qubit transition (carrier coupling) can likewise be eliminated using pulse engineering techniques with no need for changes in the optical setup or additional gates ~\cite{shapira2020}. Further work on engineering inter-qubit couplings could be directed to eliminating crosstalk interactions between pairs of neighbors, which is second order in $\eps$ but may still be significant in longer and tightly-spaced strings and under-recognized because it does not affect the fidelity of an entangled state prepared on the target ions. Another promising avenue for leveraging the techniques we have described here is to implement sets of vibrational modes specially designed to be compatible with crosstalk-insensitive qubit couplings through control of the trapping potential~\cite{teoh2021_manipulating, olsacher2020_scalable, li2025_realizing}.

\section{Acknowledgments}
We thank Kristi Beck and Brant Bowers for insightful discussions, and Ashlyn Burch and Melissa Revelle for discussions and help with implementing the experiments on QSCOUT. This material was funded in part by the U.S. Department of Energy, Office of Science, Office of Advanced Scientific Computing Research Quantum Testbed Program. This work was funded in part by the AFOSR Young Investigator Program, FA9550-24-1-0146 and NSF grant OMA-2329020. V.K. is supported by the QuICS Lanczos Graduate Fellowship. C.W. is supported through the Advancing Quantum-Enabled Technologies (AQET) traineeship program at the University of Washington.

\begin{center}
{\large \textbf{Supplemental Material}}
\end{center}

\section{SM 1: Crosstalk Insensitive Couplings with Sinusoidal Modes}

\label{supp:equispaced}

Close approximations of equispaced ion strings can be produced using trapping potentials with only harmonic and quartic components~\cite{lin2009}. The motional modes associated with strings of equispaced ions are closely-approximated by sinusoids~\cite{kyprianidis2024,johanning2016}:
\begin{align}
    b^{\text{sin}}_{m,j} = \sqrt{\frac{2-\delta_{m,1}}{N}}\cos\left( \frac{(2j-1)(m-1)\pi}{2N} \right).
    \label{eq:sinusoidal_modes}
\end{align}

These vibrational modes are more suitable for the implementation of a variety of inter-ion couplings compared to the set of modes produced in harmonic trapping potentials~\cite{kyprianidis2024}. It is proven in Ref.~\cite{kyprianidis2024} that such a collection of modes can be excited using only a global drive to produce an inter-qubit coupling that acts solely between every pair of target ions that are symmetric to each other across the center of the string, i.e. pairs of the form $(j,N+1-j)$. This coupling is in fact an instance of a crosstalk-insensitive qubit coupling because there is no coupling between any symmetric pair of target ions and their non-target neighbors. We extend this result by proving that crosstalk-insensitive couplings are possible using this set of modes for every pair of target ions that does not include the outer two ions of the string.

Describing these normal modes as $N$-element vectors defined by $b^{(m)}_j=b^\text{sin}_{m,j}$, the vectors obey the orthonormality condition $\bs{b}^{(m_1)} \cdot \bs{b}^{(m_2)} = \delta_{m_1, m_2}$. We can also consider the set of $N$-element vectors similarly defined but indexed by $m$ rather than $j$: $\tilde{b}^{(j)}_m = b^\text{sin}_{m,j}$. For $j_1,j_2\in\{1,2,\cdots,N\}$, these vectors are also orthonormal: $\bs{\tilde{b}}^{(j_1)} \cdot \bs{\tilde{b}}^{(j_2)} = \delta_{j_1,j_2}$. Extending the range of $j_1$ to be $\{1-N, 1-N+1, \cdots, 2N\}$ while keeping $j_2 \in \{1,2,\cdots,N\}$, we note that the vectors obey the modified relation $\bs{\tilde{b}}^{(j_1)} \cdot \bs{\tilde{b}}^{(j_2)} = \delta_{j_1,j_2} + \delta_{1-j_1, j_2} + \delta_{2N+1-j_1,j_2}$.

We define the motional dependence vector of the coupling between some ions $j_1$ and $j_2$ by
\begin{align}
    g^{(j_1,j_2)}_m = \tilde{b}^{(j_1)}_m \tilde{b}^{(j_2)}_m.
\end{align}
Given a set of two target ions $\mathcal{T}$ and the set of up to four nearest neighbors $\mathcal{N}$, we aim to show that $\bs{g}^{(t_1,t_2)}$ is linearly independent of the set of vectors $\mathcal{R}^{(t_1,t_2)} = \{\bs{g}^{(t,n)} : t,n \in \mathcal{T} \times \mathcal{N}\}$. To show this linear independence it suffices to identify a vector that is perpendicular to all vectors in $\mathcal{R}^{(t_1,t_2)}$ but has nonzero projection onto $\bs{g}^{(t_1,t_2)}$. We seek such a vector with the ansatz form $C^{(h)}_m = 2 \cos(h(m-1)\pi/N)$ with $h \in \{1, 2, \cdots, N-1\}$. To deduce the appropriate value for $h$, we begin by recognizing that, for $j_1<j_2 \in \{1,2,\cdots,N\}$,
\begin{align}
    C^{(h)}_m g^{(j_1,j_2)}_m = g^{(j_1+h,j_2)}_m + g^{(j_1-h,j_2)}_m.
\end{align}
Noting that $\sum_m g^{(j_1,j_2)}_m = \bs{\tilde{b}}^{(j_1)} \cdot \bs{\tilde{b}}^{(j_2)}$, we conclude that
\begin{align}
    \bs{C}^{(h)} \cdot \bs{g}^{(j_1,j_2)} & = \delta_{j_1+h,j_2} + \delta_{1-j_1-h, j_2} + \delta_{2N+1-j_1-h,j_2}
    \nonumber
    \\
    & \quad \delta_{j_1-h,j_2} + \delta_{1-j_1+h,j_2} + \delta_{2N+1-j_1+h,j_2}
    \nonumber
    \\
    & = \delta_{h, j_2-j_1} +\delta_{h, j_1+j_2-1} + \delta_{h, 2N+1-j_1-j_2}.
    \label{eq:Cdotg}
\end{align}
To arrive at the second equality above we eliminate terms that are necessarily zero due to the ranges of $j_1$, $j_2$, and $h$.

If we pick $h=t_2-t_1$ then $\bs{C}^{(t_2-t_1)} \cdot \bs{g}^{(t_1, t_2)} = 1 \neq 0$ as desired. To see that $\bs{C}^{(t_2-t_1)}$ is also orthogonal to all vectors in $\mathcal{R}^{(t_1,t_2)}$, we consider all possibilities that could lead to a nonzero value in Eq.~\ref{eq:Cdotg}: $t_2-t_1 \neq t_2-t_1 \pm 1$ for any $t_1$ and $t_2$, $t_2-t_1 = t_1+t_2-1\pm1$ only for $t_1=1$, and $t_2-t_1=2N+1-t_1-t_2\pm1$ only for $t_2=N$. Thus, as long as $t_1 \neq 1$ and $t_2 \neq N$, $\bs{C}^{(t_2-t_1)}$ is perpendicular to all vectors in $\mathcal{R}^{(t_1,t_2)}$ but not perpendicular to $\bs{g}^{(t_1,t_2)}$. We conclude that $\bs{g}^{(t_1,t_2)}$ is linearly independent of $\mathcal{R}^{(t_1,t_2)}$ for any $t_1,t_2 \in \{2,3,\cdots, N-1\}$. Choosing $\chi \propto \bs{C}^{(t_2-t_1)}$ would therefore produce a crosstalk-insensitive qubit coupling matrix.

Because $\bs{C}^{(t_2-t_1)}$ lies within $R^\perp$ and has non-zero projection onto $\bs{g}^{(t_1,t_2)}$, it provides a bound on the target-crosstalk independence parameter: $a_{t_1,t_2} := \lvert\lvert \text{proj}_{R^\perp}(\bs{g}^{(t_1,t_2)}) \rvert\rvert/\lvert\lvert \bs{g}^{(t_1,t_2)} \rvert\rvert \geq |\bs{C}^{(t_2-t_1)} \cdot \bs{g}^{(t_1,t_2)}|/[||\bs{C}^{(t_2-t_1)}||||\bs{g}^{(t_1,t_2)}||]$. Since $||\bs{C}^{(h)}||=\sqrt{2N}$ independently of $h$ and $||\bs{g}^{(t_1,t_2)}|| \leq (2/N) \sqrt{N} = 2/\sqrt{N}$, we arrive at the lower bound $a_{t_1,t_2} \geq \frac{1}{(\sqrt{2N}\sqrt{4/N})} = \frac{1}{2\sqrt{2}}$.

\section{SM 2: Multitone Entangling Pulses}

The numerical simulations of entangling gates presented in Fig.~\ref*{fig:independence_and_pulse_infidelity} in the main body of this work are for a string of 20 \Yb ions spaced equally with a separation of $3\mu\text{m}$. We consider a counterpropagating 355nm Raman beam configuration, leading to a effective wavenumber of $k=2\times 2\pi/355\text{nm}$. The beams are perpendicular to the ion string leading to motion only in the radial direction. The radial trapping frequency is taken to be $\omega_\text{rad} = 2\pi \times 4 \text{MHz}$. The set of radial modes are described by frequencies $\nu_{m}$ and ion participation vectors $b_{m,j}$. For these numerical calculations, we use the actual motional modes of finite-length equispaced ion strings, which differ slightly from the sinusoidal approximation that we analyze in SM1 due to edge effects. The Hamiltonian for the qubit states and motional modes is
\begin{align}
    \hat{H}_\text{base} = \sum_{j=1}^N \frac{\omega_0}{2} \hat{\sigma}^z_j + \sum_{m=1}^N \nu_m (\hat{a}_m^\dagger \hat{a}_m+1/2),
\end{align}
where $\omega_0$ is the energy splitting frequency of the qubit levels in each ion and $\hat{a}_m$ is the lowering operator for the $m$th mode.
Following Ref.~\cite{shapira2020}, we consider the application of a multitone laser pulse of duration $\tau_\text{gate}$. The laser has $M$ pairs of frequency components, each consisting of frequencies $\omega_0\pm\omega_l$. Both tones in a pair have the amplitude $r_l$, which is normalized such that $\sum_l r_l=1$. All ``blue" tones (positive detuning) have phase $\phi_B$ and all ``red" tones (negative detuning) have phase $\phi_R$. The peak Rabi frequency of light incident on ion $j$ is $\Omega_j$. Additionally, the phase of the tones incident on ion $j$ includes an offset of $\tilde{\phi}_j$ to account for the path length of light to the ion.
The interaction Hamiltonian is~\cite{landsman2019}
\begin{align}
    & \hat{V}(\tau) = \sum_j \Omega_j \hat{\sigma}^x_j \sum_{l=1}^M r_l [ \cos(k \hat{r}_j - (\omega_0+\omega_l) \tau + \phi_B + \tilde{\phi}_j) 
    \nonumber \\
    & \qquad \qquad + \cos(k \hat{r}_j - (\omega_0-\omega_l) \tau + \phi_R + \tilde{\phi}_j) ],
    \nonumber \\
    & = 2 \sum_j \Omega_j \cos(k \hat{r}_j - \omega_0 \tau + \phi_s + \tilde{\phi}_j) \hat{\sigma}^x_j \sum_l r_l  \cos(\omega_l \tau + \phi_m),
\end{align}
where $k$ is the wavenumber of the laser, $\hat{r}$ is the radial displacement of ion $j$ along the direction parallel to laser propagation, and $\phi_{s/m} = [\phi_R \pm \phi_B]/2$.

In the Lamb-Dicke regime $k \hat{r}_j << 1$, such that
\begin{align}
    \hat{V} \approx & \sum_j \left[ \cos(-\omega_0 \tau + \phi_s + \tilde{\phi}_j) +  k\hat{r}_j \sin(-\omega_0 \tau + \phi_s + \tilde{\phi}_j) \right]
    \nonumber \\
    & \qquad \times \Omega_j \hat{\sigma}^x_j \sum_l r_l \cos(\omega_l \tau + \phi_m).
\end{align}
The first term in parentheses describes carrier coupling that flips spin states without coupling to motion. We assume that $\omega_l$ is large enough for all $l$ such that carrier coupling is negligible and so ignore this term. If carrier coupling cannot be ignored, it is still possible to design pulses that coherently cancel it out~\cite{shapira2020}.

The radial position of each ion is expressed in terms of the motional mode positions of each ion as $\hat{r}_j = \sum_m b_{m,j} \eta_m (\hat{a}_m + \hat{a}^\dagger_m)$, where $\eta_m = k \sqrt{\hbar/2m_\text{ion}\nu_m}$ is the Lamb-Dicke parameter for ion $j$ with mass $m_\text{ion}$. The interaction Hamiltonian in the interaction picture with respect to $H_0$ becomes
\begin{align}
    & \hat{H}_\text{int}(\tau) = \sum_j \Omega_j \sum_m b_{m,j} \eta_m \sin(-\omega_0 \tau + \phi_s + \tilde{\phi}_j)
    \nonumber \\
    & \quad \times  \left[\hat{a}^\dagger_m e^{i \nu_m \tau} + \text{h.c.}\right] \left[ \hat{\sigma}^+_j e^{i \omega_0 \tau} + \text{h.c.} \right] \sum_l r_l \cos(\omega_l \tau + \phi_m),
\end{align}
where $\hat{\sigma}^+_j = \hat{\sigma}^x_j+i\hat{\sigma}^y_j$. Going forward, the rotation operator is parameterized as $\hat{\sigma}^{(\phi)}_j := -i\hat{\sigma}^+_j e^{i\phi} + \text{h.c.}$.
Under the rotating wave approximation we arrive at the general form for $\hat{H}_\text{int}$ provided in Eq.~\ref*{eq:Hint} of the main text:
\begin{align}
    \hat{H}_\text{int}(\tau) = & \frac{1}{2} \sum_j \Omega_j \sum_m b_{m,j} \eta_m \hat{\sigma}_j^{(\phi_s + \tilde{\phi}_j)} \hat{a}^\dagger_m e^{i \nu_m \tau} f_{\text{mt}}(\tau) + \text{h.c.},
\end{align}
with the multitone pulse function
\begin{align}
    f_\text{mt}(\tau) = \sum_l r_l \cos(\omega_l \tau + \phi_m).
\end{align}

The unitary describing evolution under $\hat{H}_\text{int}$ for a time $\tau_\text{gate}$ can be derived using the Magnus expansion~\cite{zhu2006_arbitrary} as 
\begin{align}
    \hat{U} = \exp(\hat{M_1}) \exp(\hat{M_2}),
\end{align}
where
\begin{align}
    & \hat{M_1} = -i \int_0^{\tau_\text{gate}} d\tau \hat{H}_\text{int}(\tau)
    \nonumber \\
    & = -\frac{i}{2} \sum_j \Omega_j \sum_m b_{m,j} \eta_m \hat{a}^\dagger_m \hat{\sigma}^{(\phi_s+\tilde{\phi}_j)}_j \int_0^{\tau_\text{gate}} d\tau e^{i \nu_m \tau} f_\text{mt}(\tau)
    \nonumber \\
    & \qquad + \text{h.c.}.,
    \\
    & \hat{M}_2 = -\frac{1}{2} \int_0^{\tau_\text{gate}} d\tau_1 \int_0^{\tau_1} d\tau_2 [\hat{H}_\text{int}(\tau_1), \hat{H}_\text{int}(\tau_2)],
    \nonumber \\
    & = -\frac{i}{4} \sum_{j_1,j_2} h_{j_1} h_{j_2} \hat{\sigma}^{(\phi_s+\tilde{\phi}_{j_1})} \hat{\sigma}^{(\phi_s+\tilde{j_2})}_{j_2} \sum_{m} \eta_m^2 b_{m,j_1} b_{m,j_2}
    \nonumber \\
    & \qquad \int_0^{\tau_\text{gate}} d\tau_1 \int_0^{\tau_1} d\tau_2 f_\text{mt}(\tau_1) f_\text{mt}(\tau_2) \sin(\nu_m [\tau_2-\tau_1]).
\end{align}

To ensure $\hat{M}_1=0$ at gate time, we require for each mode $m$ that
\begin{align}
    \int_0^{\tau_\text{gate}} \sum_{l} r_l \cos(\omega_l t + \phi_m) e^{i\nu_m t} dt =0.
\end{align}
This imposes a complex linear constraint, and so two real linear constraints, on $\bs{r}$ for each mode $m$. As long as $\bs{r}$ has more than $2N$ elements, all the constraints can be satisfied simultaneously.

For our calculations we set $\phi_B=\phi_R=0$ and so $\phi_s = \phi_m = 0$. Having chosen $\bs{r}$ to properly disentangle all modes by gate time, the gate unitary becomes $\hat{U} = \exp(\hat{M}_2)$, matching Eq.~\ref*{eq:qubit_unitary} in the main text with the rotation axis for each ion set by $\phi_j=\phi_s+\tilde{\phi}_{j} = \tilde{\phi_j}$.

\section{SM 3: Experimental Parameters}
\label{supp:experiment}

The experimental demonstration of a crosstalk-insensitive two-qubit entangling gate in a three-ion string was performed using the Quantum Scientific Computing Open User Testbed (QSCOUT) at the Sandia National Laboratory \cite{clark2021_engineering}. On the QSCOUT platform, qubits are encoded in the hyperfine ${}^2S_{\frac{1}{2}}|F=0,m_F=0\rangle \equiv |0\rangle$ and ${}^2S_{\frac{1}{2}}|F=1,m_F=0\rangle \equiv|1\rangle$ levels of \Yb ions, with a 12.6 GHz energy splitting ~\cite{sandiagaussian}. Raman transitions are used to apply single-qubit rotations and  state-dependent forces. Our experiment used a linear chain of three \Yb ions trapped in a harmonic potential with a radial trap frequency $\omega_x = 2\pi\times2.506$\,MHz and axial trap frequency $\omega_z=2\pi\times0.7$\,MHz. 

\bibliographystyle{unsrt}
\bibliography{references}

\end{document}